# An Architecture for Establishing Legal Semantic Workflows in the Context of Integrated Law Enforcement[1]

Markus STUMPTNER[a,2] Wolfgang MAYER[a], Georg GROSSMANN[a], Jixue LIU[a],
Wenhao LI[a], Pompeu CASANOVAS[bd], Louis DE KOKER[b], Danuta MENDELSON[e],
David WATTS[b], Bridget BAINBRIDGE[b]

[a] *University of South Australia, Adelaide*
[b] *La Trobe Law School, La Trobe University, Melbourne, Australia*
[d] *Autonomous University of Barcelona, Spain*
[e] *Deakin Law School, Deakin University, Melbourne, Australia*

**Abstract.** Traditionally the integration of data from multiple sources is done on an ad-hoc basis for each analysis scenario and application. This is a solution that is *inflexible*, incurs *high costs*, leads to "silos" that prevent sharing data across different agencies or tasks, and is unable to cope with the modern environment, where workflows, tasks, and priorities frequently change. Operating within the Data to Decision Cooperative Research Centre (D2D CRC), the authors are currently involved in the Integrated Law Enforcement Project, which has the goal of developing a federated data platform that will enable the execution of integrated analytics on data accessed from different external and internal sources, thereby providing effective support to an investigator or analyst working to evaluate evidence and manage lines of inquiries in the investigation. Technical solutions should also operate ethically, in compliance with the law and subject to good governance principles.

## 1. Introduction

This paper presents ongoing research of the Australian government-funded Data to Decisions Cooperative Research Centre (D2D CRC).[3] Australia's national security and law enforcement agencies are faced with a deluge of intelligence and other data. Data from sources as varied as financial transactions, immigration movements, vehicle registrations, call charge records, criminal histories, airline data, social media services, etc. results in a flood of information in disparate formats and with widely varying content. In Australia, such data is often held by individual federal, state or territory agencies and inter-agency access to and sharing of data is generally subject to multiple laws and complicated rules and agreements [20][21]. Accessing relevant data as well as linking and integrating them in a correct and consistent way remains a pressing challenge, particularly when underlying data structures and access methods change over time. In addition to this challenge, a large volume of data needs to be handled. Usually only a fraction of current volumes can be analyzed. The Big Data challenge is to extract maximum value from this flood of data through the use of smart analytics and machine enablement.

---

[1] Pre-print. A previous version of this paper was presented at the Third Workshop on Legal Knowledge and the Semantic Web (LK&SW-2016), EKAW-2016, November 19th, Bologna, Italy. This version has been submitted for publication at AICOL-2017.
[2] Corresponding Author.
[3] http://www.d2dcrc.com.au/

Traditionally the integration of data from multiple sources is done on an ad-hoc basis for each analytical scenario and application. This is a solution that is *inflexible*, *costly*, entrenches "silos" that prevent sharing of results across different agencies or tasks, and is unable to cope with the modern environment, where workflows, tasks, and priorities frequently change. Working within the D2D CRC, one group of authors of this article are currently involved in the Integrated Law Enforcement Project, which has the goal of developing a federated data platform to enable the execution of integrated analytics on data accessed from different external and internal sources, in order to provide effective support to an investigator or analyst working to evaluate evidence and manage lines of inquiry in the investigation. This will be achieved by applying *foundational semantic technologies* based on the *meta-modelling* of data models and software systems that permit alignment and translation by use of *model-driven transformations* between the different APIs, services, process models and meta-data representation schemes that are relied upon by the various stakeholders. It will also provide easily adapted interfaces to third party data sources currently outside of the stakeholders' reach, such as financial transactions. The other group of authors are involved in the D2D CRC's Law and Policy Program, which aims to identify and respond to the legal and policy issues that arise in relation to the use of Big Data solutions by Australian law enforcement and national security agencies.

A 2015 systematic ACM review and mapping [1] of papers on online data mining technology intended for use by law enforcement agencies identified eight main problems being addressed in the literature: (i) financial crime, (ii) cybercrime, (iii) criminal threats or harassment, (iv) police intelligence, (v) crimes against children, (vi) criminal or otherwise links to extremism and terrorism, (vii) identification of online individuals in criminal contexts, and (viii) identification of individuals. The survey also included an array of technologies capable of application to Open Source Intelligence (OSINT), i.e. data collected from publicly available sources in the fight against organized crime and terrorism: Artificial Intelligence, Data Fusion, Data Mining, Information Fusion, Natural Language Processing, Machine Learning, Social Network Analysis, and Text Mining.

Data integration in this context raises serious legal compliance and governance challenges. While the *Onlife* Manifesto considers the use of self-enforcing technologies as the exception, or a last resort option, for coping with the impact of the information revolution [2], nothing prevents the regulation of OSINT in accordance with existing legislation and case law, international customary law, policies, technical protocols, and best practices [3]. Indeed, compliance with existing laws and principles is a pre-condition for the whole process of integration, as information acquisition, sharing and analysis must occur within the framework of the rule of law.

We have taken this complex set of issues into account in our paper on architecture and information workflows. In order to foster trust between citizens and national security and law enforcement agencies, a commitment to transparency and respect for privacy must be preserved. However, addressing these issues in practice is difficult; in order to achieve a good outcome a more nuanced approach may be required. For example, an insistence upon 'full transparency' may not be desirable for citizens and law enforcement agencies alike if it undermines operational secrecy. Rather, the goal is to identify an outcome that maintains public accountability, understanding that to do so requires effort. The identification of relevant legal, regulatory and policy requirements is the starting point of this process.

## 2. Architectural Challenges

*2.1. Data Integration and Matching*

Many prototypes for data matching exist [4]. Matching systems rely either on hand-crafted rules or use simple lexical similarity and concept tree based similarity measures. Complex data structures and entire Service API interface specifications are not covered. Besides extensions for complex structures, simplification of human input and incremental match maintenance are open issues for further research.

Mapping of relational data sources to semantic models is still a predominantly manual activity. Standards, such as XML-DB and RDB2RDF can represent only syntactic mappings. Academic tools (e.g., Karma) allow mapping of relational sources to rich semantic models based on past mappings. Incremental match maintenance (if the model on either side evolves) and support for query APIs and meta-data attributes are not supported in the current tools.

Linked data uses standards such as RDF and OWL for linking knowledge sources in the Web. Although links can be established manually or with the help of various application- and source-specific algorithms, dealing with the semantic interpretation of links spanning multiple sources, the integration of data models, meta-data, and the possible unintended consequences of linking entities is often left to application programmers.

NIEM[4] has emerged as a standard for information exchange between government agencies in the U.S.A. The standard specifies data models for specific message exchanges (in XML format) between two endpoints, and covers core data elements that are commonly understood and defined across domains (e.g. 'person', 'location') as well as community specific elements that align with individual domains (e.g. immigration, emergency management, screening). However, the standard is weak in relation to meta-data and provenance information, and security considerations are orthogonal. Moreover, the architecture is designed for enterprise application integration, not Big Data analytic interfaces. There is no equivalent standard in Europe yet although the EU General Data Protection Regulation (GDPR) (2016/679), which comes into force in May 2018, is motivating its development [5].

Current research focuses on schema mapping where a relationship between data specifications is established [4][6]] in a semi-automated way and requires a domain ontology. Current challenges include dealing with the evolution of schema- and data structure, for example, how to re-establish links across data sources that have been changed, integrating data on the record level – not only on the schema level, and integration of data streams which is particularly important for integration of events.

---

[4] National Information Standard Model, https://www.niem.gov/

*2.2. Meta-data*

Meta-data management is addressed in various proprietary ways in most commercial databases, intelligence tools, and Big Data platforms. A federated meta-data mechanism is required that spans multiple vendor tools and can capture and manage meta-data such as provenance, data quality, and linkage information at the right granularity (attribute/fact level) for a policing and intelligence context.

Linking data and data access processes to related legal policies and workflows is required but often not provided explicitly. Although there are a growing number of databases that use licenses (CC), most of them do not contain any reference to licenses [7]. Yet, there are some research attempts to compose them [8] or to facilitate their use within a copyright term bank [10], or through a general framework [9].

Meta-data approaches for linked data platforms, such as Resource Description Framework (RDF) annotations, are not standardized and possess no widely agreed-upon semantics. The W3C is currently working on security standards for linked data.[5] However, this standard will be generic and may not meet the specific needs of intelligence and policing applications (e.g. it will lack the capacity to establish and preserve the provenance of the meta-data, which is critical when dealing with data that is sourced across governmental or organisational boundaries). Temporal aspects and the degree of confidence in meta-data are also not considered.

Information governance is as important in Big Data initiatives as in traditional information management projects. Gartner has identified information governance as one of the top three challenges of Big Data analytics initiatives.[6] SAS is also singling out information governance and data quality as major challenges to the success of analytics projects.[7]

In traditional information management initiatives, the focus has been on absolute control of the data attributes such as accuracy, consistency, completeness and other data quality dimensions. Initiatives such as meta-data management and master data management have assisted in creating 'single versions of the truth' for sharing information assets [33].

Big Data initiatives involving the placement of disparate data sets into 'data lakes' for analysis have significant information governance issues as they have the potential to force knowledge of contextual awareness and semantic understanding. Once analytical models have been created their operationalization will be restricted without data curation and lineage metadata.

---

[5] https://www.w3.org/Metadata/
[6] Gartner Data & Analytics Summit 2017. 20 – 21 February / Hilton Sydney.
[7] SAS Institute (Suite of Analytics Software), esp. Best Practices in Enterprise Data Governance, https://www.sas.com

*2.3. Workflow Orchestration*

Commercial tools predominantly rely on proprietary implementations of analytic tool chains and workflows. Although there is some support for exchange of analytical processes (e.g., through an UIMA specification) in commercial tools (Leidos TeraText) tools are confined to a federation comprising a single vendor's analytic tool chain in many cases. Some platforms, e.g. SAS IM, can access Hadoop file systems and generate analytic scripts for such Map-Reduce architectures. Big Data frameworks (Hadoop, YARN, RabbitMQ) support efficient and flexible data pipelines but require configuration and custom code for integration with vendor tools for each individual application.

Scientific workflow tools (Kepler, Taverna) provide process design and limited execution monitoring and (coarse-grained) provenance mechanisms. However, extensions and novel meta-data models are required to suit the specific tools for policing/analytics and fine-grained provenance.

If more than a few selected data sources are accessed in a data pipeline, data access itself can become an issue since end-user analysts cannot be expected to know all of the underlying technical data models and languages used by the individual systems (e.g., (Geo)SPARQL, (Geo)JSON, XML, etc.). Effective end-user analytic processes therefore must abstract from source-specific query languages at the user level. There is ample academic work on query translation, but mostly for relational databases [11].

Existing analytic tools/products assume that all data is available locally, and require (at least partial) ETL/ingest and indexing of data to facilitate linking, matching, search, deduplication, and quality checking. If federated cross-platform analytic processes are to be supported, systematic mechanisms are required for managing cached and replicated data across vendor tool chains, and local updates made in an intelligence tool should be fed back into the federation.

Multiple workflows and policies may apply to analysis tasks (in particular to identity resolution), individual cases and agencies, and workflows may differ based on availability of data, timeline, and security credentials. As such, in addition to executing a given analytic process it is also desirable to suggest and configure suitable analytic processes specific to the analyst's situation.

## 3. Architectural Overview

The overall architecture for an Integrated Law Enforcement System is described in Figure 1. The architecture comprises 8 broad categories of services depicted as blue rectangles.

The chosen approach is consistent with the TOGAF Architecture Development Method [28], in that requirements management within the project is ongoing and changing stakeholder requirements, changes in partnerships, and technological innovation result in updates to the specifics. With regard to managing strategies according to the approaches listed in [27], the setting of the project can be categorised as a bottom up approach (due to the independence of different agencies), open specifications, and a focus on technical integration and a reference architecture approach. Centralisation efforts that may lead to a more top-down approach are underway but at this point not clear in terms of their impact.

The project has defined an open architecture for data/meta-data management and analytic processes. As such it will translate the best practices from Enterprise Application Integration to the "Big Data" analytic pipelines. The work addresses aspects related to data and metadata modelling and storage, modelling and execution of analytic processes across multiple analytic tools and data sources. Central to this architecture is a framework for effective semi-interactive entity linking and querying of linked data. The project has been working on a comprehensive data management framework that relies on a well-defined shared data and meta-data model supported by vendor-agnostic interfaces for data access and execution of processes comprising analytic services offered by different tools. In the following sections we discuss the main elements of the architecture.

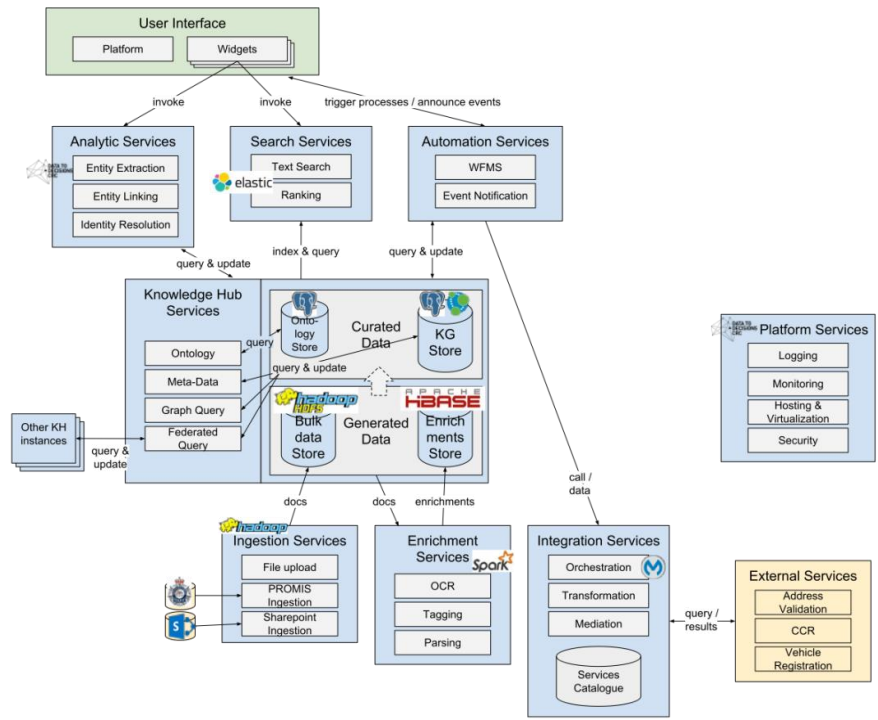

**Figure 1.** Architecture overview

## 3.1. Knowledge Hub Services

The central *Knowledge Hub Services* area comprises the data stores and related data management services. It governs the repository of data held in the node of the federated architecture and exposes data and the schema via query services to front-end interfaces.

*3.1.1. Data Stores*

The Knowledge Hub's data stores implement a polyglot architecture comprising of multiple technologies tailored to different categories of data. It is partitioned into the Curated Data area which includes databases that hold information that has been confirmed by a user, and the Generated Data area which hold data that have been derived by automatic enrichment processes but not yet confirmed.

The Knowledge Graph store holds the collection of linked entities. It holds facts and meta-data about entities and their links whose veracity has been confirmed. Data in this store is predominantly structured, linked, and associated with meta-data, providing a semantic entry point for operating on the base data. The information is represented in terms of an Ontology describing the main entity types, relation types, their attributes, and meta-data attributes. The ontology itself is also represented explicitly and can be queried. Due to differing access rules and access patterns for data/meta-data and ontology, entities/links/meta-data are stored separately from the Ontology. The linked data store implements a directory of entities and links enriched with appropriate meta-data and source information such that detailed information can be obtained from authoritative sources that may be external to the system. This approach is needed as data in the law enforcement domain is dispersed among a number of systems owned and operated by different agencies. As such no centrally controlled database can feasibly be put in place in the foreseeable future.

The Bulk Data store holds documents and binary objects (e.g., videos), and the enrichments store holds derived information that may facilitate analysis and can be promoted into the curated data area. As curated and generated data exhibit different lifecycles, volume, and access patterns, these are held in separate data stores.

*3.1.2. Ontology and Meta-Data*

The contents of the data stores are governed by an Ontology that describes the domain-specific taxonomy of entity types, their properties, and relationships. Akin to a schema definition in a relational database, the ontology acts as a reference for knowledge organization and aids in the integration of information stemming from external sources, where it acts as a reference for linking and translating information into a form suitable for the knowledge hub. The adoption of a linked data approach supported by an ontology provides the flexibility needed to evolve data and schema over time and supports semantic interpretation of the linked elements.

The ontology has been designed specifically for the law enforcement domain and includes a broad spectrum of data types, including structured entities such as persons, organizations, vehicles, communication events, and their relationships; and unstructured data including text documents, video and audio recordings. The ILE ontology is too large to reproduce it in full in this paper; it comprises 19 high-level domain concepts which are further refined into a total of ~140 concepts and a taxonomy of ~400 specialized relationship types. The full ontology has been documented in [23]. These domain concepts are closely aligned with the draft National Police Information Model (NPIM), complemented with relevant aspects drawn from the NIEM standard[8] and concepts related to case management. The provenance model is an extension of PROV-O [24].

---

[8] https://www.niem.gov/

The ontology is complemented by a meta-data ontology that defines the meta-data attributes that are associated with each entry in the knowledge hub. Meta-data information includes information about provenance, links to entries in external systems and document store, temporal qualification, security and access control descriptors, and information about acquisition process and modification events. Meta-data is one of the cornerstones of information management in the knowledge hub, as the process and timing of information acquisition must be documented meticulously in order to satisfy the legal requirements related to evidence collection. Meta-data elements can be attached to each element (entity, property, link). This relatively fine-grained approach has been selected to be able to support entity linking and merging of information stemming from multiple external sources. The resulting data and meta-data model serve as the foundation for information use, governance, data quality protocols, analytic pipelines, exploration and justification of the results.

*3.1.3. Query Services*

The services exposed by the Knowledge Hub component relate to querying, keyword search, and meta-data access. At the time of writing, the implementation of the architecture supports structured and keyword search queries about related entities. Graph matching and graph analytics are not currently implemented directly in the hub and must be performed in external tools.

The architecture embraces location- and representation transparency principles that enable front-end applications to access data irrespective of *where* the data is stored and rely on a uniform representation defined by an ontology irrespective of *how* the data is stored in a source system. If a knowledge hub node cannot satisfy a query based on the information it holds, queries to external system can be spawned to acquire the requested information. For this purpose, each knowledge hub maintains a registry of sources that can be queried for information. Adapters for each source type abstract from the specific technologies and message formats used to access the sources. The results obtained from other systems are then expressed in the common representation and merged into a consolidated list of results using entity linking and ranking algorithms.

*3.2. Ingestion and Integration Services*

Ingestion services provide functions for importing (bulk) data from external systems, such as legacy case management systems and document repositories. The ingestion process follows a pipeline architectural style where data is processed and enriched in stages: data is acquired from the source system, the content is indexed, entities are extracted, annotated with meta-data and provenance information, and linked to relevant existing information in the knowledge hub. The results are represented in terms of the ontology and added to the knowledge hub.

The heterogeneity in representation and data format among different sources presents challenges related to information interpretation and transformation into the ontology used within the data platform. Declarative data transformation methods are employed to convert the different external representations into a common data model and link the resulting structure to the ontology governing the knowledge hub. Ontology matching techniques [11] are used in a semi-interactive process to match and convert user supplied data, and graph mining and matching techniques have been developed to improve the mapping of implicit relationships discovered between extracted entities. The mapping from proprietary data representations to the representation used in the knowledge hub is performed via declarative transformation specified in the Epsilon Transformation Language framework [25]. This approach enables flexible configuration of the transformation rules as sources are added. Moreover, the explicit representation of the transformation facilitates analysis of impact of changes as the internal ontology evolves.

Interoperability with existing data sources and systems can be achieved by constructing executable mappings from the (meta-)data model to the individual system's data models and APIs. Our work goes beyond existing Extract-Transform-Load (ETL) and data access approaches in that the mapping will facilitate bi-directional communication to allow for propagating updates to/from the federated knowledge hub, and model-driven mapping technologies will facilitate maintenance of schema transformations in case source systems undergo extensions or data format changes. We intend to rely on proven meta-modelling techniques for early detection and semi-automated resolution of mapping problems at the interfaces to legacy systems. This approach will help avoid problems related to failing ETL processes and subtle issues arising from changing data sources. Currently, these issues are predominantly left for manual resolution by software engineers.

The pursued approach is well-established in Enterprise Application integration but has only recently been considered in the form of "Big Service" integrated pipelines [12], where a shared architecture comprising of an integrated shared data model and implementation-agnostic service APIs is described. This project aims to translate this idea to the policing and intelligence domain where Big Data requirements are prevalent. Given that the number of sources relevant to policing and intelligence has been increasing, unless systematic data management and access mechanisms are implemented, data quality, provenance and maintenance issues are likely to worsen if the current siloed approach is continued.

*3.3. Analytic Services*

Analytic services related to the data platform include entity extraction from unstructured text [26], entity linking, similarity calculation and ranking. Entity linking and ranking methods that are effective for sparse data are being developed. Moreover, the inclusion of meta-data attributes in the resolution and ranking calculations is a distinguishing feature of this work. Services provided by commercial tools, such as network analysis and entity linking/resolution solutions, can be integrated in the modular architecture.

The project will provide an efficient, open orchestration platform for Big Data applications that can incorporate capabilities from multiple tools made available through services. Integration of multiple COTS tools will furthermore enable analysts to use familiar languages and tools for exploring information and semi-interactive linking/resolution while relying on capabilities beyond their single tool to perform search, linking, analysis tasks. For static information models, specific-purpose tool chains could be coded by software engineers; however, for end-user specified analytic pipelines and evolving linked data models (e.g. case-specific taxonomies) a comprehensive data and processing architecture is needed.

For efficiency of certain analytic tasks, data may need to be replicated and indexed in multiple locations, subject to security and data access considerations. Keeping these cached copies consistent is a non-trivial problem that we will address through leveraging incremental data propagation and eventual consistency mechanisms. This capability will benefit end users who can rely on automation of complex data management processes, avoid batch ETL jobs and error-prone manual coding of data/meta-data transformation pipelines, and improve data consistency and meta-data capture.

*3.4. Automation Services*

Automation services provide workflow and process orchestration functions. Workflow services will facilitate the enactment of work processes such as acquiring authorization and warrants. Supported by a set of user-facing "widgets" and a library of tasks and processes, common standard processes can be planned and automatically triggered and executed. Process standardization for common tasks may improve efficiency and compliance with policies and legal requirements. In the domain of law enforcement, some processes vary depending on the context of the investigation. A task ontology for investigation planning could capture the semantics of key process steps as well as machine-interpretable descriptions of the roles of various actors and data objects, e.g. evidence, in the investigation process. Automatic configuration of workflows and tracking of their execution may reduce the potential for errors. Moreover, appropriate provenance and chain of evidence can be established by linking the information obtained in the course of an investigation with the process activity and timeline that led to its acquisition. At the time of writing, the automation services component is not yet implemented in software.

*3.5. Security*

Information security is one of the main concerns in a system designed for law enforcement. Trust in the sharing platform is paramount as an absence of trust and security protocols will prevent sharing of most data. Any data sharing platform in this domain needs to be capable of operating in a multi-agency environment where each agency may have its own security and information sharing policies and protocols. It is difficult to envision a single system and access control policy that would simultaneously satisfy all stakeholders' requirements.

The approach taken in the architecture presented here rests on two principles: (i) a fine-grained security model where access privileges can be associated with each individual fact that in the knowledge hub (akin to the Accumulo database management systems), and (ii) a federated network of linked knowledge hubs that collaborate to provide access to information. The granular access privilege model enables precise control of what information can be disclosed (e.g. some attributes and relationships associated with selected entities may be classified or restricted whereas others may be accessible to all authorized users). The federated architecture aims to build trust in the sharing platform by maintaining control of data access within each individual source organization. Queries are dispatched to multiple nodes and executed under the local node's access policy. The results are then transmitted and collated at the originating node where a query was posed. At the time of writing, the precise access control model and full implementation of the access federation remain the subject of future work.

*3.6. Legal Workflow Processing*

Another key concern is the incorporation of constraints into the workflow execution to ensure compliance with laws, policies and procedures, including agency legislation and applicable privacy requirements. Natural Language parsing can be used to elicit event specifications that could then be translated into business rules in an executable formal language and issued to an event processor in the knowledge hub [19]. These rules would be used to check and guarantee conformance of analytic processes/workflows and data usage.

At present, there exists a substantive body of works on law and semantic languages —LegalXML, LegalRuleML, RDF and legal ontologies in OWL modelling interoperability and reasoning [13]. However, in the field of policing, law enforcement and security, research and experience have shown that this relationship is by no means simple. Law and policies are subject to contextual and dynamic interpretation across different jurisdictions, legal systems and policy environments [22]. Legal requirements cannot be comprehensively itemized/programmed, nor can the rule of law, privacy and data protection be hard-coded, in particular, where judgement or an exercise of discretion is required [14] [15]. Thus, indirect strategies [16] and design tactics [17], including privacy engineering and risk management, will need to be employed to ensure both compliance with the Australian legal system, and beyond that, the proactive adoption of privacy and data protection safeguards and ethical principles. [18].

*3.7. Example*

The system has not been tested with end-users users, but the following provides an example illustrating its application and the legal challenges faced by the designers. In the context of police investigations, work processes can be supported by partially automated workflows that help investigators carry out activities efficiently while, at the same time, ensuring that each activity is linked to appropriate supporting information. For example, the planning underpinning an application for a search warrant of premises by an officer of the Australian Federal Police in terms of the *Crimes Act 1914* (Cth) could be partly automated.

The common law imposed significant restrictions on the use of search and seizure powers by government officials and constables based on the inviolability of property interests:

> *Against that background, the enactment of conditions, which must be fulfilled before a search warrant can be lawfully issued and executed, is to be seen as a reflection of the legislature's concern to give a measure of protection to those interests. To insist on strict compliance with the statutory conditions governing the issue of a search warrant is simply to give effect to the purpose of the legislation.[9]*

The law therefore provides a range of control measures to protect the rights of individuals affected or potentially affected by a search warrant. Apart from the procedures prescribed by the *Crimes Act 1914*, outlined in this example, other Acts may also be relevant, for example the *Australian Federal Police Act 1979* (Cth) and the *Privacy Act 1988* (Cth). There would also be internal agency procedures to be followed, including the Commonwealth Director of Public Prosecution's *Search Warrant Manual* for obtaining and executing warrants under Commonwealth law. A range of legal questions may therefore arise and these may differ from case to case.

---

[9] *George v Rockett* (1990) 170 CLR 104 at pp 110-111

Section 3E of the *Crimes Act 2014* sets out a number of requirements that must be met before a valid search warrant can be issued. In broad terms, a successful search warrant application involves two steps. The first is to assemble the necessary material necessary to enable the applicant to present sufficiently persuasive material, on oath or affirmation, to enable an 'issuing officer' to be satisfied 'that there are reasonable grounds for suspecting that there is, or there will be within the next 72 hours, any evidential material at the premises.'[10] The second is for the issuing officer, once so satisfied, to address the requirements set out in section 3E(5) in the warrant itself. These include a description of the offence to which the warrant relates,[11] a description of the premises to which the warrant relates[12] and the kinds of evidential material to be searched for.[13]

The background processes that are needed to support compliance with requirements such as these could be automated using a federated data architecture. Investigation planning could be supported by an ontology describing goals and activities that may be conducted in the course of an investigation. Each goal would be associated with supporting sub-goals and activities as well as information requirements. If an element is added to the investigation plan, subordinate elements could be automatically added and, where possible, executed automatically. This would require a careful preparatory work, as each line of investigation may rise separated legal issues.

For example, if an investigator adds a line of inquiry (e.g. representing the criminal history, if any, of the owner of the premises) to the investigation plan in the case management system, the integrated information architecture as described in this paper would automatically enable a search of its knowledge hub for relevant information, including the criminal history of the subject, whether the person owns a registered firearm and, where relevant, whether previous applications were made for warrants relating to the same person or premises, and their outcomes. The system would then populate an application for an arrest warrant, which, after being sworn or affirmed by the applicant, could be presented, potentially automatically, to the issuing officer and the applicant would be informed of the outcome. If the outcome is positive, execution of the warrant must be planned. To facilitate risk mitigation, the system could further determine whether there is information indicating that other persons reside at the same address, whether they pose a potential threat, or whether there are any potential threats linked to surrounding premises and their occupants that should be considered. Such information may be obtained from data sources within law enforcement, such as a case management system containing structured person records, documents and notes; and from sources external to the agency, such as council rate bills, electoral roll information, and information extracted from social media posts by the subject. Linking this information to the corresponding elements in the investigation plan facilitates a more comprehensive and efficient consideration of available information and automation facilitates appropriate execution of mandated investigation practices.

---

[10] S 3E(1) of the *Crimes Act 1914*
[11] S 3E(5)(a) of the *Crimes Act 1914*
[12] S 3E(5)(b) of the *Crimes Act 1914*
[13] S 3E(5)(c) of the *Crimes Act 1914*

Many legal questions arise in relation to the design of an effective automated system, for example: What are the different types of information that an investigator would wish to access? Where relevant information is highly sensitive to another investigation conducted by a different team, is the investigator entitled to access that information? Once collected for purposes of the warrant, can personal information of the occupants and residents in the area be used and stored for future investigations? Where answers to these legal questions may be clear, they may differ across Australian states and territories [29] [30], and the architecture should be flexible enough to accommodate all legal requirements. In this project we are therefore addressing two separate but linked sets of problems: (i) the coexistence of both artificial and human decision-making and information processes; and (ii) the modelling of specific legal requirements arising from different legal and government sources [31] [32]. We are considering a blended "RegTech" perspective[14] to be applied to law enforcement and security with the due legal protections in place.

---

[14] http://www.investopedia.com/terms/r/regtech.asp

## 4. Intention and future work

This paper outlined the data management architecture for supporting law enforcement agencies under development at the D2D CRC. Although some of it is confidential at a granular level, the architectural overview, meta-data driven integration, and legal workflow processing can be disclosed for academic and scientific discussion. We have shown that extensible domain ontologies and semantic meta-data are an essential pillar of long-term data management in a domain where the variety of data and complex analytical processes are dominant. In the law enforcement domain, work processes, mandated procedures-, approval- and data acquisition processes are just as important as the collected information. Moreover, in the age of advanced data analytics, discoveries are increasingly based on automated collection and analysis of data. As a result, questions related to the way in which these processes are conducted and under what circumstances their results may be used are increasingly important and must be considered simultaneously with the design of the business processes and supporting information systems. We contend that policy models for law enforcement and national security purposes should and can be based on an appropriate understanding and implementation of all relevant legal requirements. We intend to further explore the specific requirements pertaining to policing and the wider law enforcement context in Australia and devise appropriate models and implementations that can support and govern the information sharing activities conducted within and across organizational boundaries.